\documentclass[12pt,preprint]{aastex}

\newcommand{\kms}          {\mbox{${\rm km~s^{-1}}$}}

\newcommand{\ee}           {\mbox{$^{-2}$}}

\def\simgt{\lower.5ex\hbox{$\; \buildrel > \over \sim \;$}}
\def\simlt{\lower.5ex\hbox{$\; \buildrel < \over \sim \;$}}

\def\Nstar{\mbox{$N_*$}}
\def\Sig3{\mbox{$\Sigma_{\rm 3 Myr}$}}
\def\Nc{\mbox{$N_{\rm c}$}}

\def\PN{\mbox{$P_{N_*}$}}
\def\PMN{\mbox{$P_{M_{\rm SN}|N_*}$}}
\def\PslrN{\mbox{$P_{\rm SLR|N_*}$}}
\def\Pslr{\mbox{$P_{\rm SLR}$}}
\def\PNslr{\mbox{$P_{\rm N_*|SLR}$}}
\def\fsn{\mbox{$f_{\rm SN}$}}
\def\fdisk{\mbox{$f_{\rm d}$}}
\def\barfdisk{\mbox{$\bar f_{\rm d}$}}
\def\tdisk{\mbox{$t_{\rm d}$}}
\def\Msn{\mbox{$M_{\rm SN}$}}
\def\tsn{\mbox{$t_{\rm SN}$}}
\def\tsf{\mbox{$t_{\rm sf}$}}
\def\rc{\mbox{$r_{\rm c}$}}
\def\re{\mbox{$r_{\rm e}$}}
\def\rs{\mbox{$r_{\rm s}$}}
\def\fe{\mbox{$f_{\rm e}$}}
\def\vexp{\mbox{$v_{\rm exp}$}}

\begin{document}

\title{On the likelihood of supernova enrichment of protoplanetary disks}
\author{Jonathan P. Williams$^{1,3}$}
\author{Eric Gaidos$^{2,3}$}
\affil{$^1$ Institute for Astronomy, University of Hawaii,
2680 Woodlawn Drive, Honolulu, HI 96822}
\affil{$^2$ Department of Geology and Geophysics, University of Hawaii,
1680 East-West Road, Honolulu, HI 96822}
\affil{$^3$ NASA Astrobiology Institute, University of Hawaii}
\email{jpw@ifa.hawaii.edu, gaidos@hawaii.edu}
\shorttitle{Supernova enrichment of protoplanetary disks}
\shortauthors{Williams \& Gaidos}

\begin{abstract}
We estimate the likelihood of direct injection of supernova ejecta
into protoplanetary disks using a model
in which the number of stars with disks decreases linearly with time,
and clusters expand linearly with time such that their
surface density is independent of stellar number.
The similarity of disk dissipation and main sequence lifetimes implies
that the typical supernova progenitor is very massive, $\sim 75-100\,M_\odot$.
Such massive stars are found only in clusters with $\simgt 10^4$ members.
Moreover, there is only a small region around a supernova within which
disks can survive the blast yet be enriched to the level observed
in the Solar System. These two factors limit the
overall likelihood of supernova enrichment of a protoplanetary
disk to $\simlt 1$\%. If the presence of short lived
radionucleides in meteorites is to be explained in this way, however,
the Solar System most likely formed in one of the largest clusters in
the Galaxy, more than two orders of magnitude greater than Orion,
where multiple supernovae impacted many disks in a short period of time.
\end{abstract}

\section{Introduction}
Meteoritic evidence for live $^{60}$Fe, a short lived radionucleide (SLR),
in the early Solar System strongly suggests that a supernova occurred
shortly before the formation of the planets \citep{Tachibana06}.
30 years ago, and in the context of another SLR, $^{26}$Al,
\cite{Cameron77} proposed that supernova ejecta may be incorporated
into planetesimals via the triggered collapse of the pre-solar nebula.
More recently, \cite{Hester05} have suggested that SLRs are directly
injected into newly formed protoplanetary disks.

The evolution of young stellar clusters and circumstellar disks
has been well studied \citep{Lada03,Haisch01}.
This allows us to quantify the likelihood that a supernova will
occur close enough in time and space to a planet-forming disk to
provide the inferred abundances of SLR in the Solar System.
In this Letter, we address two questions:
what is the likelihood of SLR enrichment of a disk by a supernova
and what cluster properties maximize this likelihood?

We summarize the relevant information on cluster and disk
properties that we use in our model in \S\ref{sec.parameters}.
We find an analytic solution for the enrichment likelihood in a cluster
in \S\ref{sec.starburst} under the simplest assumption of a starburst,
generalize to a finite formation period with a Monte Carlo
simulation in \S\ref{sec.finite}, and consider the effect
of multiple supernovae in \S\ref{sec.multiple}.
We calculate the overall enrichment probability and determine
the cluster size that maximizes disk enrichment in \S\ref{sec.implications}.

\section{Parameters of the problem}
\label{sec.parameters}
\subsection{Cluster number distribution}
We adopt a cluster number distribution,
$d\Nc/d\Nstar\propto \Nstar^{-2}$,
that is consistent with both
young, embedded clusters $\simlt 3$\,Myr \citep{Lada03},
and older, optically visible open clusters \citep{Elmegreen97}.
The minimum cluster size is largely a matter of semantics for our
calculations since small clusters are exceedingly unlikely
to have supernovae within the maximum disk lifetime.
The maximum cluster size, determined from the radio measurements of
the ionizing luminosity of HII regions, is $N_{\rm *,max}=5\times 10^5$
\citep{McKee97}.

There are proportionally more chances to find a star in a larger group 
so the differential probability that a star is found within a cluster
with \Nstar\ members, $dP/d\Nstar\propto 1/\Nstar$.
We will plot probabilities as a function of the logarithm of the
cluster size and define
$\PN \equiv dP/d\ln\,\Nstar\,\Delta\ln\,\Nstar = P_0\Delta\ln\,\Nstar.$
The constant, $P_0$, is determined by the
condition that the integral of \PN\ over all clusters be
equal to the total probability that a star is born in a cluster.
\cite{Lada03} estimate that between $70-90\%$
of all stars form in clusters and that 90\% of these form in
clusters with $\Nstar\geq N_{\rm *,min}=100$.
Taking an average of 80\% for the former gives
$P_0=0.8\times 0.9/\ln(N_{\rm *,max}/N_{\rm *,min})=0.085$.

\subsection{Cluster expansion}
\cite{Lada03} show that the number of detectable
clusters declines with age and estimate that only about 10\% of clusters
survive as recognizable entities beyond 10\,Myr.
The spatial dispersion of nearby moving groups such as the
TW Hya and $\beta$\,Pic
associations show how quickly stars migrate away from their siblings.
For the greater Scorpius-Centaurus region,
with a spatial scale of 200\,pc and age $\sim 30$\,Myr \citep{Zuckerman04},
the implied expansion speed is $\sim 3$\,\kms.

All clusters at a given age, regardless of stellar number,
have a similar average stellar surface density, $\Sigma_*=\Nstar/\pi \rc^2$,
where \rc\ is the cluster radius \citep{Adams06}.
For the list of small clusters, $\Nstar\sim 30-1000$,
in \cite{Lada03}, the average surface density is
$\Sigma_*\simeq 100$\,stars\,pc\ee\ at an average age $t\sim 3$\,Myr.
The clusters in \cite{Carpenter00}, defined from 2MASS data,
have lower surface densities, $\Sigma_*\simeq 30$\,stars\,pc\ee.
For massive star forming regions, \cite{McKee03} find a much
higher characteristic surface density, $1\,{\rm g\,cm}^{-2}$,
corresponding to $\Sigma_*\sim 10^4$\,stars\,pc\ee\
but in much younger objects, $t\ll 1$\,Myr.
We assume a constant expansion speed such that $\rc=\vexp t$
and $\Sigma_*\propto t^{-2}$, so the equivalent stellar surface
density at 3\,Myr is at least one, and possibly as many as two,
orders of magnitude less. To bracket the possibilities, we consider
the values, $\Sig3=30, 100$, and 300\,stars\,pc\ee.

\subsection{Stellar mass function}
\label{sec.imf}
The stellar initial mass function (IMF) appears remarkably
uniform over a range of cluster sizes \citep{Kroupa01}.
We use the
Scalo mass function for which the number of stars with masses,
$M>8\,M_\odot$, that will become core collapse supernovae
are a fraction $\fsn=3\times 10^{-3}$ of all stars and follow
a power law distribution, $d\Nstar/dM_*\propto M^{-(1+\alpha)}$,
with index $\alpha=1.5$ \citep{Scalo86}.
The cumulative distribution of supernova progenitors in a cluster
containing \Nstar\ stars is therefore
\begin{equation}
\label{eqn.cumulative}
\Nstar(>M_*)=\fsn\Nstar\frac{(M_u/M_*)^\alpha-1}{(M_u/M_l)^\alpha-1}
\end{equation}
where the lower and upper limits of the progenitor distribution are
taken to be $M_l=8\,M_\odot\leq M_*\leq M_u=150\,M_\odot$.
The existence of an upper limit is clear, although its
actual value can only be statistically estimated \citep{Figer05}.

\subsection{Disk lifetimes and supernova timescales}
Mid-infrared surveys of stellar clusters,
for which an average age can be determined, show that the fraction
of stars with disks, \fdisk, is observed to decrease from unity
at $< 1$\,Myr to zero at $\tdisk=6$\,Myr \citep{Haisch01}.
The decrease in disk fraction to zero is approximately linear with time,
\begin{equation}
\fdisk(t)=
\begin{cases}
1-t/\tdisk & t\leq\tdisk,\\
0          & t>\tdisk.
\end{cases}
\label{eqn.fdisk}
\end{equation}
UV radiation from O stars can rapidly photoevaporate
the outer radii of protoplanetary disks \citep[e.g.][]{Storzer99}
but sufficient mass to form planetary systems may remain
bound to the star \citep{Williams05,Eisner06}.
Several of the clusters in the Haisch et al. survey and other
similar studies \citep[e.g.][]{Mamajek04}
contain O stars and the average disk fraction does not appear
to be adversely affected, at least for luminosities comparable to Orion.
Note also that whatever the star formation scenario, whether instantaneous,
gradual, or induced, it is effectively incorporated into this
empirical formalism.

For supernova timescales,
we use an empirical fit to the \cite{Schaller92} stellar evolution models,
$\log_{10}\,\tsn = 1.4/(\log_{10}\,\Msn)^{1.5}$,
where \tsn\ is in Myr and \Msn\ is in solar masses.

If all stars in a cluster are coeval, the similarity of circumstellar
disk and massive star lifetimes implies that no more than about half
the disks remain when the first supernova occurs, even for the
most massive progenitors. The least massive star that could explode
within 6\,Myr is $\Msn=30\,M_\odot$,
and is only likely to be found in clusters with $\Nstar\simgt 2600$.
If clusters are formed more gradually, some disks may exist at later
times and slightly lower mass progenitors may play a role (\S3.2).

\subsection{Proximity to supernova blast}
Disk enrichment places tight constraints on spatial scales too.
A minimum mass solar nebula (MMSN) disk with radius 100\,AU
around a solar mass star will be stripped by supernova ejecta
within 0.2\,pc \citep{Chevalier00}.
Matching the abundances of the ejecta with the meteoritic record,
however, requires that disks lie within
0.22\,pc for a $25\,M_\odot$ progenitor \citep{Looney06} and
0.3\,pc for a $40\,M_\odot$ progenitor \citep{Ouellette05}.
We find that even larger masses are more likely sources of enrichment
due to their shorter main sequence lifetimes.
These can enrich a larger volume, out to $\sim 0.4$\,pc for a
$100\,M_\odot$ progenitor.
We therefore consider a radial range, $\rs<r<\rs+\re$,
where $\rs=\re=0.2$\,pc within which disks can survive
the supernova and be enriched to the level observed in the Solar System.
The number of stars in this ``enrichment zone''
depends on the cluster density profile and size.

Observations of Orion \citep{Hillenbrand98}
and other clusters \citep[e.g.][]{Muench03}
show that the number of stars per unit area declines approximately
inversely with angular distance from the center
(albeit with some significant substructure in some cases).
The inferred stellar volume density profiles are therefore
approximately inverse square, $n_*\propto 1/r^2$,
and the number of stars increases linearly with radius,
$\Nstar(<r)=\Nstar(r/\rc)$, where $\rc$ is the cluster radius.
The most massive stars in a cluster are generally found near
its center \citep{Lada03}.
Assuming that this is the case for the supernova progenitor,
the fraction of stars in the enrichment zone is
\begin{equation}
\label{eqn.Ne}
\fe(t)=\frac{\Nstar(<\rs+\re)-\Nstar(<\rs)}{\Nstar}
      =\frac{\re}{\rc}.
\end{equation}

The stellar motions in a cluster have characteristic value,
$\vexp=\rc/t=(\Nstar/9\pi \Sig3)^{1/2}$.
For example, a typical cluster in \cite{Lada03} with $\Nstar=10^3$,
$\Sig3=100$\,pc\ee, has $\vexp=0.6$\,\kms\ but the velocities
are higher in larger clusters with similar surface densities.
To estimate the fraction of stars within the enrichment zone,
we assume that the cluster maintains its inverse square density profile,
and, when considering the effect of multiple supernovae (\S\ref{sec.multiple}),
that stars move independently through this zone.

\section{Calculation of the enrichment likelihood}
\subsection{Cluster formation in a starburst}
\label{sec.starburst}
Under the assumption that all the stars in a cluster form at the same
time, i.e. in a starburst, the first supernova will be the most massive
star and the enrichment likelihood can be calculated analytically.

The conditional probability that the most massive star
in a cluster
has mass \Msn\
is the expected number of stars of this mass times the probability
that there are none more massive \citep[see][]{Williams97},
\begin{equation}
\PMN=\frac{\alpha\fsn\Nstar}{[(M_u/M_l)^\alpha-1]M_u}
     \left(\frac{M_u}{\Msn}\right)^{1+\alpha}\,e^{-{\cal N}_*(>M_{\rm SN})}.
\end{equation}

If we further assume that the disk lifetime is independent of cluster
location then the fraction of disks that exist within the enrichment
zone can be separately factored as $\fdisk\fe$.
Evaluating this at the time of the supernova and integrating over
all possible progenitors gives the likelihood that a disk in a cluster of a
given size is enriched with SLR at the level observed in our Solar System,
\begin{equation}
\PslrN=\int_{M_l}^{M_u}\fdisk(\tsn)\,\fe(\tsn)\,\PMN\,dM_{\rm SN}.
\end{equation}

Figure~\ref{fig.pslr_single} shows that \PslrN\ has a similar form
independent of \Sig3\ with a broad maximum centered on
$\Nstar\sim 8000$. Small clusters are unlikely to have supernovae
within the disk lifetime and large clusters expand more rapidly
so relatively few stars lie within the enrichment zone.
The absolute likelihood increases for higher surface
densities since the expansion speed will be lower, and the
enrichment fraction higher, for a given cluster number.

\subsection{Cluster formation over a finite duration}
\label{sec.finite}
The starburst assumption is a simplification:
young clusters contain protostars in a range of evolutionary states
suggesting that they are built up over a period $\tsf\sim 1$\,Myr
\citep{Lada03}.

For a prescribed star formation rate (SFR), $\dot N_*(t)$,
the average disk fraction in the cluster is
$\barfdisk(t)=\int_0^{t}\fdisk(t-t')\,\dot N_*(t')\,dt'/N_*.$
The time dependence of the SFR in clusters is unknown.
The simplest assumption is that it is constant,
$\dot N_*=\Nstar/\tsf$. In this case, for $t>\tsf$,
\begin{equation}
\barfdisk(t)=
\begin{cases}
1-(t-\tsf/2)/\tdisk             & t\leq\tdisk,\\
(\tsf+\tdisk-t)^2/2\tdisk \tsf  & \tdisk<t\leq\tdisk+\tsf\\
0                               & t>\tdisk+\tsf.
\end{cases}
\end{equation}

We proceed by randomly sampling a star from the IMF
at each time interval $\Delta t=1/\dot N_*$. The birth time
is added to the main sequence lifetime for each star with
$M_*>8\,M_\odot$ to determine when the first supernova occurs.
At this time, $t_1$, \PslrN\ is calculated as the product of the
disk fraction above times the enrichment fraction, where the
cluster size $\rc=\vexp t_1=(N_*/9\pi\Sig3)^{1/2}t_1$\,pc.
As $t_1$ is a stochastic variable, we average
over $10^5$ simulations for each cluster number.

Figure~\ref{fig.pslr_single} shows that \PslrN\ is relatively insensitive
to \tsf, and much more dependent on \Sig3.
Peak probabilities are slightly lower for a finite formation time as
the higher disk fraction at the time of the supernova is offset by the
larger cluster size and lower enrichment fraction
The extended period over which a supernova can impact circumstellar
disks allows for lower mass progenitors, however.
The average supernova mass is $74\,M_\odot$ for $\tsf=1$\,Myr
compared to $98\,M_\odot$ for the starburst.
In each case, such high masses are favored because of the short
disk lifetimes and are only likely to be found in very large
clusters, $\Nstar\simgt 10^4$.

\section{Multiple supernovae}
\label{sec.multiple}
Large clusters, $\Nstar\simgt 5000$, should have more than one
supernovae within the maximum disk lifetime, $\tdisk+\tsf$.
We model the effect of additional supernovae by summing enrichment probabilities
using the same Monte Carlo formulation as in \S\ref{sec.finite}.
The assumption here is that cluster dynamics move new star-disk systems 
into the enrichment zone by the time of the next supernova.
Two relatively small corrections are made: first, the
fraction of possible disks that may be enriched is decreased by
the fraction that are stripped by the previous supernova, $\rs/\rc$;
second, the fraction of disks in the enrichment zone, \fe,
is decreased by $\vexp\Delta\tsn/\re$ if this is less than 1,
where $\Delta\tsn$ is the interval between supernovae, to allow
for migration of new disks into this region. In practice, only
at most a few percent of disks are close enough to a supernova
to be destroyed and, except for the very largest clusters,
the interval between supernovae is large enough that the enrichment
zone is continually refreshed.

The results are shown in Figure~\ref{fig.pslr_multiple}
both for starburst and extended SFR scenarios, $\tsf=0,3$\,Myr.
The enrichment likelihood continually increases to the largest
clusters where many tens of supernovae can contribute.
The inset shows the total number of supernovae within $\tdisk+\tsf$
and the number that enrich half of the overall total for a fiducial
$\tsf=1$\,Myr.

\section{Implications}
\label{sec.implications}
\begin{figure}[t]
\vskip -0.25in
\includegraphics[width=5in,angle=90]{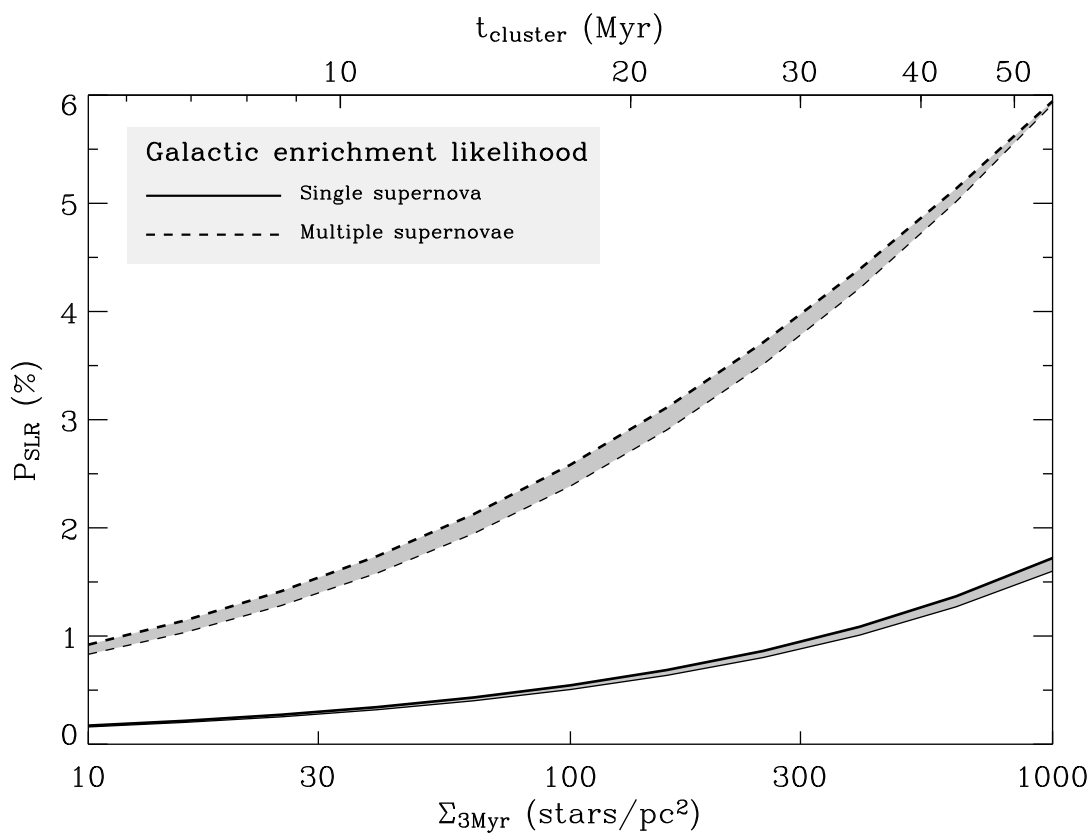}
\caption{The dependence of the overall likelihood of SLR enrichment,
summed over all cluster sizes and allowing for stars not formed in
clusters, as a function of the average stellar surface density at 3\,Myr.
The variation with cluster formation time from $\tsf=0-3$\,Myr
is indicated by the shaded region with the starburst scenario shown
as the thicker line.}
\label{fig.pslr_sigma}
\end{figure}

The overall probability of supernova enrichment for {\em any} star
is determined by integrating the enrichment likelihood,
or conditional probability of enrichment given cluster size,
over the cluster number distribution,
$\Pslr=\int_{\ln\,N_{\rm *,min}}^{\ln\,N_{\rm *,max}}\PslrN\,P_0\,d\ln\,\Nstar.$
This ``Galactic enrichment likelihood'' is plotted versus a wider
range of average surface densities in Figure~\ref{fig.pslr_sigma}.
The incorporation of the cluster number distribution adds some
uncertainty to the absolute numbers
but the general form, in particular the strong dependency on
\Sig3\ and near-independence on \tsf, is inherited from \PslrN\ and is robust.
As the surface density decreases with time, the cluster will
become harder to identify. A rough estimate of
the cluster lifetime, $t_{\rm c}$,
for which $\Sigma(t_{\rm c})=3$\,stars\,pc\ee,
comparable to the field star density, is shown on the upper axis.
Given that $\sim 90$\% of all clusters do not survive beyond 10\,Myr,
and that planetary systems are disrupted in longer lived systems
\citep{Adams01},
we conclude that supernova enrichment of protostellar disks is a
highly unlikely event, affecting $\simlt 1$\% of all stars in the Galaxy.

\begin{figure}[t]
\vskip -0.25in
\includegraphics[width=5in,angle=90]{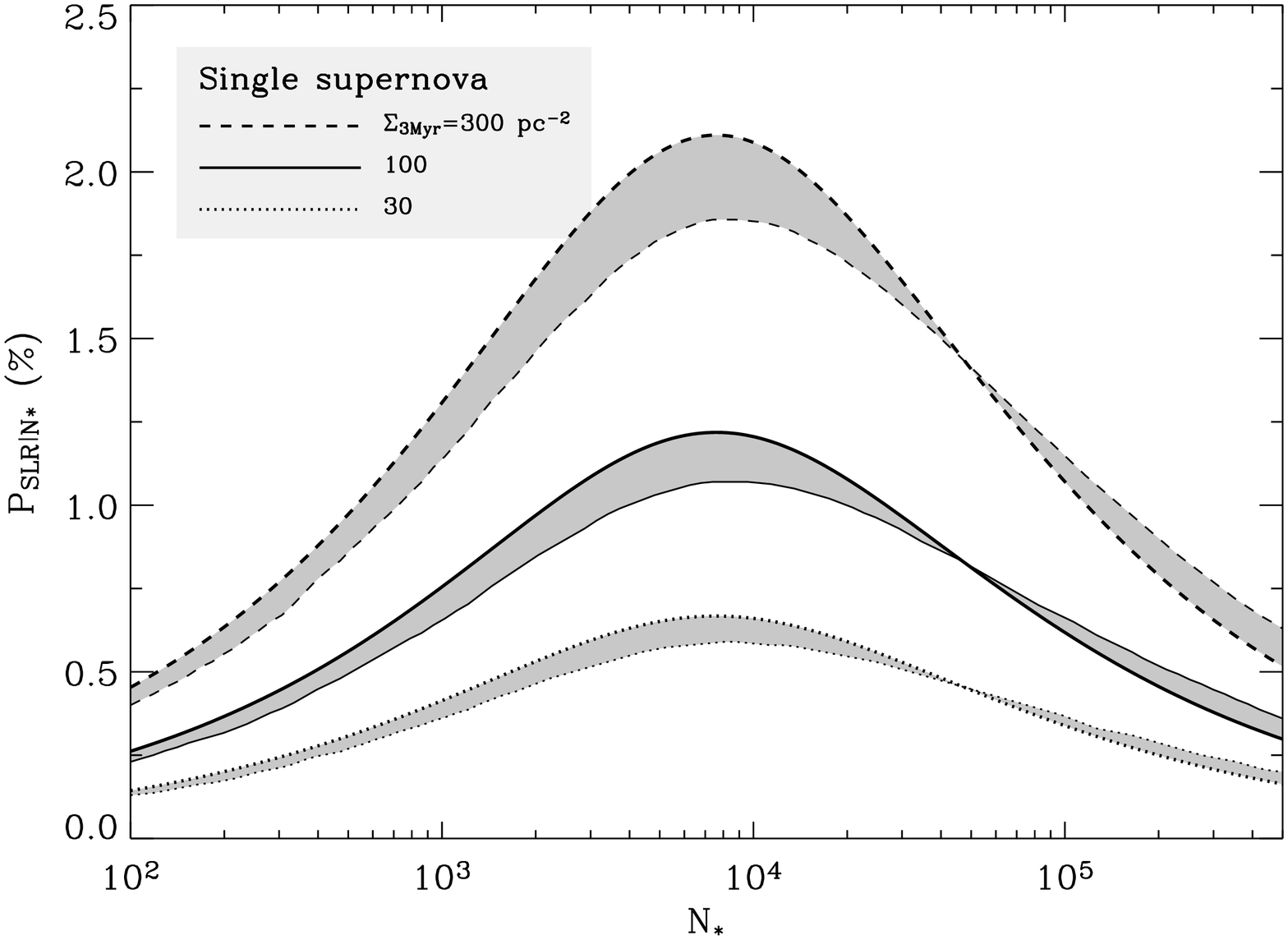}
\caption{The enrichment likelihood from a single supernova event
versus stellar number for three values of the average surface density
at 3\,Myr.
The shaded regions show the variation with cluster formation time
from $\tsf=0-3$\,Myr, with the starburst scenario indicated by the
thicker line.}
\label{fig.pslr_single}
\end{figure}

\begin{figure}[t]
\vskip -0.25in
\includegraphics[width=5in,angle=90]{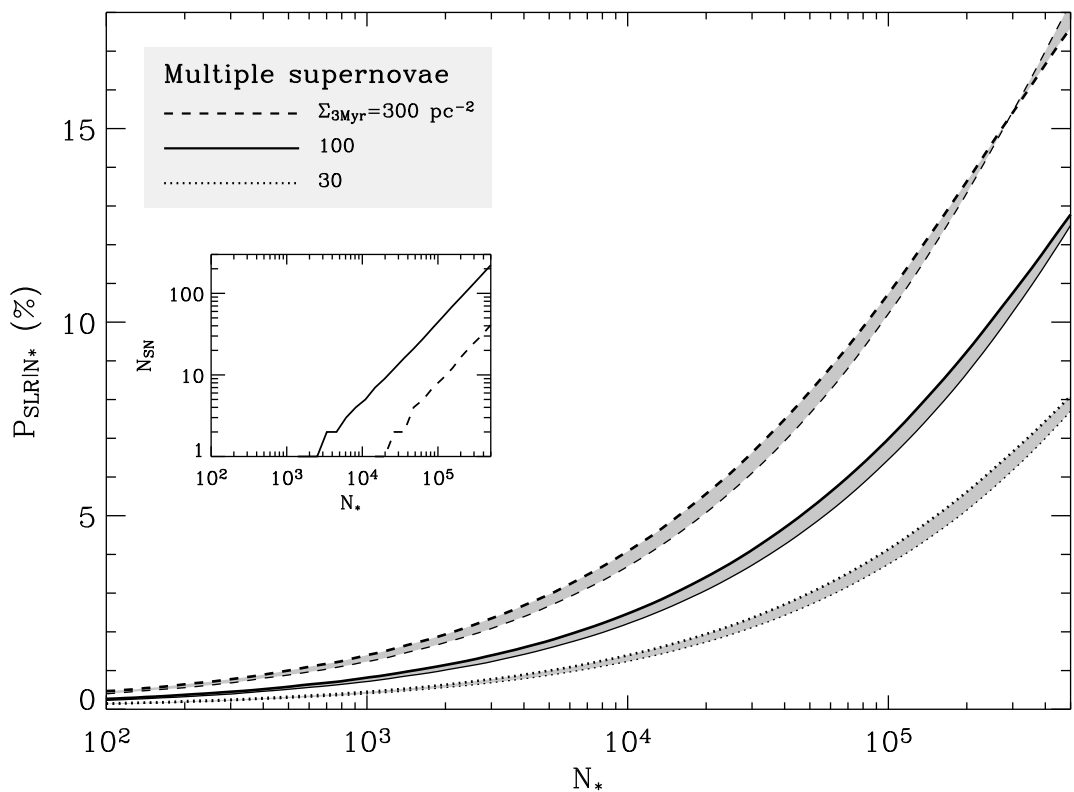}
\caption{The enrichment likelihood as a function of stellar number,
as in Figure~\ref{fig.pslr_single}, but for the case of multiple supernovae.
The inset shows the total number of supernovae (solid line) per cluster
and the number at which half the final enrichment fraction was reached
(dashed line) for the case $\tsf=1$\,Myr.}
\label{fig.pslr_multiple}
\end{figure}

How, then, to explain the presence of $^{60}$Fe in the early Solar System?
If, however unlikely, it was injected into the protoplanetary disk
from a nearby supernova, then a Bayesian estimate of the most likely
cluster size is $\PNslr=\PslrN\PN/\Pslr$.
Because \PN\ is independent of \Nstar\ and \Pslr\ is a normalization
factor, the conditional probability of stellar number given enrichment
is directly proportional to the enrichment likelihood plotted in
Figures~\ref{fig.pslr_single}, \ref{fig.pslr_multiple}.
This strongly favors our Solar System's origin in the largest clusters
in the Galaxy, exemplified by NGC\,3603 \citep{Moffat94},
more than two orders of magnitude more luminous than Orion,
where multiple supernovae can potentially enrich many disks.

A key assumption in this conclusion is that the disk fraction is
independent of cluster location. Recent work by \cite{Balog07},
however, shows that \fdisk\ is a factor $2-3$ lower than
equation\,\ref{eqn.fdisk} in the central 0.5\,pc of NGC\,2244,
a cluster intermediate in luminosity between Orion and NGC\,3603.
This would decrease \PslrN\ by the same amount, lessen or even
nullify the increase toward large \Nstar\ in Figure~\ref{fig.pslr_multiple},
and only strengthen the conclusion that the direct injection of
supernova ejecta into a protoplanetary disk is a very unlikely event.
Moreover, if one or more massive stars is the source of other SLRs,
particularly $^{26}$Al, in the early Solar System, then their implied
scarcity in other planetary systems may have important implications for
the thermal history of planetesimals in those systems \citep{Ghosh98}.

\acknowledgments
This work was motivated by engaging converstaions with Sasha Krot and Gary Huss
and is supported by the NASA Astrobiology Institute under
Cooperative Agreement No. NNA04CC08A.
We also thank Fred Adams for comments.

\clearpage

\end{document}